\documentclass[aps,prd,twocolumn,superscriptaddress,nofootinbib]{revtex4-2}

\usepackage{graphicx}
\usepackage{xspace}
\usepackage{xcolor}
\usepackage{hyperref}
\usepackage{booktabs}
\usepackage{amssymb}
\usepackage{siunitx}
\usepackage{amsmath}
\newcommand{\dcc}{LIGO-P1900360\xspace}

\newcommand{\aligo}{Advanced LIGO\xspace}

\newcommand{\gw}[1][]{gravitational wave#1 (GW#1)\renewcommand{\gw}[1][]{GW##1\xspace}\xspace}
\newcommand{\bbh}[1][]{binary black hole#1 (BBH#1)\renewcommand{\bbh}[1][]{BBH##1\xspace}\xspace}
\newcommand{\bns}[1][]{binary neutron star#1 (BNS#1)\renewcommand{\bns}[1][]{BNS##1\xspace}\xspace}
\newcommand{\kde}[1][]{kernel density estimator#1 (KDE#1)\renewcommand{\kde}[1][]{KDE##1\xspace}\xspace}
\newcommand{\snr}[1][]{signal-to-noise ratio#1 (SNR#1)\renewcommand{\snr}[1][]{SNR##1\xspace}\xspace}
\newcommand{\map}{maximum-posterior (MaP)\renewcommand{\map}{MaP\xspace}\xspace}
\newcommand{\fap}[1][]{false-alarm probability#1 (FAP#1)\renewcommand{\fap}[1][]{FAP##1\xspace}\xspace}
\newcommand{\pval}[1][]{$p$-value#1\xspace}
\newcommand{\psd}[1][]{power spectral density#1 (PSD#1)\renewcommand{\psd}[1][]{PSD##1\xspace}\xspace}

\newcommand{\mnras}[0]{Mon.\ Not.\ R.\ Astron.\ Soc.}

\newcommand{\aap}[0]{Astron. \& Astrophys.}

\newcommand{\supplement}{appendix\xspace} 

\newcommand{\code}[1][]{\texttt{#1}\xspace}
\newcommand{\pycbc}{\code{PyCBC}\xspace}
\newcommand{\pycbcinf}{\code{PyCBC\,Inference}\xspace}

\newcommand{\dynesty}{\code{dynesty}\xspace}

\newcommand{\Msun}{M_\odot}

\newcommand{\unlens}{\mathrm{U}}
\newcommand{\lens}{\mathrm{L}}

\newcommand{\bayes}{\mathcal{B}}
\newcommand{\blu}{\ensuremath{\bayes_{\lens/\unlens}}\xspace}

\newcommand{\evi}{\mathcal{Z}}

\newcommand{\ZU}{\evi_\unlens}
\newcommand{\ZL}[1][]{\evi_{\lens#1}}
\newcommand{\murel}{\mu_\mathrm{rel}}
\newcommand{\delay}{\Delta t}

\begin{document}

\preprint{LIGO-P1900360}

\title{Search for strongly lensed counterpart images of binary black hole mergers\\ in the first two LIGO observing runs}

\newcommand{\icg}{\affiliation{University of Portsmouth, Institute of Cosmology and Gravitation, Portsmouth PO1 3FX, United Kingdom}}
\author{Connor McIsaac}
\email[]{connor.mcisaac@port.ac.uk}
\icg
\author{David Keitel}
\email[]{david.keitel@ligo.org}
\icg
\affiliation{Departament de F\'isica, Universitat de les Illes Balears, IAC3 -- IEEC, Crta. Valldemossa km 7.5, E-07122 Palma, Spain}
\author{Thomas Collett}
\icg
\author{Ian Harry}
\icg
\author{Simone Mozzon}
\icg
\author{Oliver Edy}
\icg
\author{David Bacon}
\icg

\date{14th October 2020}

\begin{abstract}
Strong gravitational lensing can produce multiple images of the same gravitational-wave signal,
each arriving at different times and with different magnification.
Previous work has explored if lensed pairs exist among the known high-significance
events from the LIGO and Virgo collaboration's GWTC-1 catalogue and found
no evidence of this.
However, the possibility remains that weaker counterparts of these events are present in the data,
unrecovered by previous searches.
We conduct a targeted search specifically looking for sub-threshold lensed images of
known \bbh observations from GWTC-1.
We recover candidates matching three of the additional events first reported by Venumadhav et al.~(2019),
but find no evidence for additional \bbh events.
We also find no evidence that any of the Venumadhav et al.~observations are lensed counterparts.
We demonstrate how this type of counterpart search
can constrain hypotheses about the overall source and lens populations and
we rule out at very high confidence the extreme hypothesis
that all heavy \bbh detections are in fact lensed systems at high redshift
with intrinsic masses $<15\Msun$.
\end{abstract}


\maketitle

\section{\label{sec:intro}Introduction}

Dense accumulations of matter, such as galaxies or galaxy clusters,
can bend the path of light from sources behind them,
an effect known as gravitational lensing~\cite{Schneider:1992}.
Similarly, \gw[s] can be lensed
by masses between the source and observer
(see e.g. ~\cite{Wang:1996as,Oguri:2019fix}).
In the strong lensing regime, multiple images are produced with a delay between arrival times.

Since 2015,
\aligo and Virgo~\cite{TheVirgo:2014hva,TheLIGOScientific:2014jea}
are regularly detecting \gw[s]~\cite{LIGOScientific:2018mvr,gracedb}
from coalescing \bbh[s] and \bns[s].
Such a signal is described by a set of parameters including the masses, spins,
location and orientation of the source.
The observed (detector-frame) masses are increased by cosmological redshift~\cite{Krolak:1987ofj}.
For lensing of \gw[s] with wavelengths in the \aligo band
by galaxy or cluster lenses,
geometric optics apply~\cite{Bontz:1981aps,Takahashi:2003ix}.
This means that for multiple images of the same event,
the lensed waveforms will be identical up to
different arrival times, amplitudes and phases~\cite{Dai:2017huk},
with apparent positions on the sky that are indistinguishable to \gw observatories.

The official LIGO-Virgo event catalog GWTC-1~\cite{LIGOScientific:2018mvr}
from the O1 and O2 observing runs
contains coalescences of 10 \bbh[s] and one \bns.
The possibility that some of these are
lensed images of a single event
has previously been suggested~\cite{Smith:2017mqu,Broadhurst:2018saj,Oguri:2018muv,Broadhurst:2019ijv}.
A systematic study~\cite{Hannuksela:2019kle} found no evidence
for multiple images, or any other lensing effects,
among the 10 \bbh[s].

However, one observational signature of strong lensing has not yet been systematically tested.
Large relative magnification between images of the same event can
lead to ``subthreshold'' counterparts \cite{Li:2019osa} of the known events,
which the broad searches conducted for GWTC-1 were not able to confidently extract from the data.
In this paper, we perform 10 separate targeted reanalyses of \aligo O1 and O2 data~\cite{Abbott:2019ebz,gwosc:O1,gwosc:O2}
searching for faint lensed counterparts.

We describe the general setup of our sub-threshold targeted searches
and subsequent candidate validation
in Sec.~\ref{sec:setup}.
This method can also readily form the basis for robust searches for lensed counterparts
in future observing runs.
We then present our results in Sec.~\ref{sec:candidates},
finding no evidence for previously unknown \bbh candidates above background,
but recovering some of the non-GWTC-1 events previously reported in \cite{Venumadhav:2019lyq,Nitz:2019hdf}.
Additional Bayesian inference demonstrates that these are more likely to be independent mergers
with parameters coincidentally similar to GWTC-1 events than actual lensed counterparts.

While the expected rate of strongly lensed events at current sensitivity is very low \cite{Ng:2017yiu,Oguri:2018muv},
even our null results can provide relevant constraints on the astrophysical population of lensed \bbh sources.
As a demonstration of this, in Sec.~\ref{sec:uls} we rule out the hypothesis of
\cite{Broadhurst:2018saj} who proposed that most, if not all, of the
heavy \bbh observations in GWTC-1
could be strongly lensed images of \bbh mergers with component masses below 15 solar masses.

\section{\label{sec:setup}Search setup}

The problem we wish to solve is the following. We have a confident \gw observation that has already been identified
in the data. We want to observe lensed counterpart images to that observation. Such counterpart images, if they
exist, would be identical to the original image except that they would be shifted in time, have a different amplitude
and potentially have a shifted coalescence phase~\cite{Dai:2017huk}. In addition to potential counterpart images
the data will also contain additional \gw signals from other sources, including weak signals not
identified with previous searches. The problem, therefore, is not only to identify potential lensed counterpart
images in the data, but also to distinguish between lensed counterpart images and unrelated \gw signals from other mergers.

The optimal method to identify lensed counterpart images of a \gw observation
would be to perform Bayesian inference on the full dataset~\cite{Ashton:2019wvo}.
This would need to include prior knowledge
of the rate and properties of counterpart images given the information known about the \gw signal.
It would also need to include prior information about the population properties
of any other, unrelated \gw signals.
One would then attempt
to determine the posterior probability of a lensed counterpart image, or images, being present in the data.
However, performing such a search over significant stretches of an observation run
would have a prohibitively high computational cost,
and the choice of population priors would require significant modelling work.

We therefore pursue a more pragmatic approach, based
on current standard search methods to identify isolated \gw signals from compact binary mergers.
Most of these methods use matched filtering over a grid of template waveforms,
using a set of assumptions and analytical maximization to restrict the search parameter space to that of the
component masses and component spins~\cite{Allen:2005fk,Babak:2012zx,Usman:2015kfa,Messick:2016aqy}.
Building on this standard approach, we perform our search for lensed counterpart images in two steps.
In our initial step we use the \pycbc search method~\cite{Usman:2015kfa,pycbc1.11.14} to search for compact binary mergers
using only a single search template waveform.
This allows us to extract weak candidates from the data that match well with that single template,
such as the lensed counterparts that we target.
However, given that LIGO and Virgo have observed numerous \bbh[s] in a relatively small region of
parameter space~\cite{LIGOScientific:2018mvr,LIGOScientific:2018jsj},
it is also possible that our search will observe additional \gw
signals from separate sources. We therefore perform a second step where,
assuming a particular candidate observed in the first step is astrophysical,
we use fully coherent Bayesian inference~\cite{Biwer:2018osg} to
compute the posterior probability between
the hypothesis that it is a lensed image of the original event
and the hypothesis that it came from a separate source.
These steps are described in more detail in the rest of this section.

\subsection{\label{sec:data_selec}Data selection}

For each \bbh event from GWTC-1,
we search the entire observing run in which it was found.
Here we assume that the break between O1 and O2 (316 days)
is longer than any time delays expected from typical astrophysical lenses --
typically less than $\sim$1 month~\cite{Li:2018prc}.
The total two-detector coincident time of
publicly available data~\cite{Abbott:2019ebz,gwosc:O1,gwosc:O2}
from the two LIGO detectors~\cite{TheLIGOScientific:2014jea}
is 48 days for O1 and 117 days for O2.
These durations already exclude any stretches of problematic data quality using the data quality categories discussed in \cite{Abbott:2019ebz}
along the same criteria as laid out in \cite{Usman:2015kfa} and used in the GWTC-1 \pycbc search.
Similar to the standard \pycbc search used for GWTC-1~\cite{LIGOScientific:2018mvr,DalCanton:2017ala},
we do not use Virgo~\cite{TheVirgo:2014hva} data for the initial search stage,
though we use it for the Bayesian inference step
on some events in the last month of O2.

We cannot include the short data stretch around GW170608
when the LIGO Hanford detector was nominally out of observing mode~\cite{Abbott:2017gyy}:
For robust statements on any candidates from this stretch,
we would have needed to include and characterize all data from other periods
when one or both of the detectors were in similar states
to produce a consistent background,
and data from such periods is not public.
Instead, we accept this short stretch as just another blind period,
similar to any other when there was no data in nominal observing mode
from both detectors at the same time.

\subsection{\label{sec:templates}Template selection}

We perform a separate search for lensed images of each \bbh from GWTC-1.
We do not include the \bns event GW170817 in this analysis,
since its close distance and extensive electromagnetic observational coverage~\cite{GBM:2017lvd}
already rule out strong lensing.
Each search uses a single aligned-spin template waveform,
generated using the SEOBNRv4\_ROM model~\cite{Purrer:2014fza,Purrer:2015tud,Bohe:2016gbl}.

To select the template parameters
we have started from the public posterior samples~\cite{gwosc:GWTC-1}
produced with the precessing IMRPhenomPv2 waveform~\cite{Hannam:2013oca,Husa:2015iqa,Khan:2015jqa}.
Since there is no evidence for precession in any of these events~\cite{LIGOScientific:2018mvr}
and the \pycbc search pipeline in its standard configuration
currently relies on aligned-spin waveforms,
we select the aligned-spin subset of parameters from the peak of each posterior.
To do so,
we obtain a four-dimensional Gaussian \kde~\cite{scikit-learn} in
$\{m_1,m_2,a_{1z},a_{2z}\}$
on each set of samples
and determine the \map set of parameters from it.
Here, $m_1$ and $m_2$ are the \bbh['s] component masses in the detector frame
while $a_{1z}$ and $a_{2z}$ are the spin magnitudes along the binary's orbital angular momentum axis.
We then generate a SEOBNRv4\_ROM waveform
at each set of \map parameters to serve as our search template.
The parameters of these 10 templates are listed in Table~\ref{table:searches}.

\begin{table}[t]
\resizebox{\columnwidth}{!}{
\renewcommand{\arraystretch}{1.25}
\begin{tabular}{crrrrrr}
\toprule
event    & $m_{1}$ [$\Msun$] & $m_{2}$ [$\Msun$] & $a_{1z}$ & $a_{2z}$ & $d_\mathrm{post}$ [Mpc] & $d_\mathrm{single}$ [Mpc] \\
\midrule
GW150914 & 37.46             & 34.47             & -0.02    & -0.02    & $1634\substack{+4 \\ -4}$   &     $1752\substack{+12 \\ -13}$ \\
GW151012 & 25.13             & 17.89             &  0.00    &  0.00    & $923\substack{+4 \\ -4}$    &     $1048\substack{+14 \\ -14}$ \\
GW151226 & 14.72             &  8.57             &  0.26    &  0.04    & $562\substack{+2 \\ -2}$    &      $604\substack{+7 \\ -7}$   \\
GW170104 & 37.45             & 23.79             & -0.03    & -0.03    & $1342\substack{+5 \\ -5}$   &     $1433\substack{+16 \\ -16}$ \\
GW170608 & 10.55             &  9.01             &  0.01    &  0.01    & $643\substack{+2 \\ -2}$    &      $675\substack{+7 \\ -7}$ \\
GW170729 & 79.45             & 48.50             &  0.60    &  0.05    & $2782\substack{+10 \\ -10}$ &     $2965\substack{+29 \\ -29}$ \\
GW170809 & 41.27             & 28.01             &  0.03    &  0.03    & $1548\substack{+8 \\ -9}$   &     $1602\substack{+15 \\ -15}$ \\
GW170814 & 32.48             & 29.42             &  0.06    &  0.03    & $1536\substack{+4 \\ -4}$   &     $1580\substack{+11 \\ -12}$ \\
GW170818 & 40.62             & 34.79             & -0.06    & -0.05    & $1604\substack{+5 \\ -5}$   &     $1757\substack{+16 \\ -16}$ \\
GW170823 & 49.06             & 40.75             &  0.03    &  0.03    & $2013\substack{+8 \\ -8}$   &     $2076\substack{+25 \\ -25}$ \\
\bottomrule
\end{tabular}
}
 \caption{\label{table:searches}
  Configurations and sensitivity metrics of the 10 searches
  for counterparts of GWTC-1 \bbh events.
  The first four columns after the event name give the parameters
  of the aligned-spin template used in each search.
  (Masses in the detector frame.)
  $d_\mathrm{post}$ is the sensitive distance of the search,
  determined by using injections drawn from the full posterior.
  $d_\mathrm{single}$ is the sensitive distance when using only the
  search template parameters for injections.
  Sensitive distances here are not re-interpreted to account for lensing magnification.
  }
\end{table}

Our choice to only use a single template to search for lensed counterpart
images for each GWTC-1 observation
is very different from the approach presented by \cite{Li:2019osa}
who reduced the original GWTC-1 template bank by a more modest factor.
To validate our choice, we compute the matches
\begin{equation}
 m(h_1,h_2) =
 \max\limits_{t_0,\phi_0} \left( \frac{h_1}{\sqrt{(h_1|h_1)}} \left| \frac{h_2}{\sqrt{(h_2|h_2)}} \right. \right) \,,
\end{equation}
defined using the usual inner product $(h_1|h_2)$ \cite{Allen:2005fk}
which depends on the detector noise curve,
between our (aligned) \map waveforms
and any draws from the whole set of (precessing) posterior samples.
This corresponds to the fraction of the \snr that we expect to recover under ideal conditions
when searching for signals drawn from the full posterior using the single template.
We compute these matches separately for each detector,
estimating its \psd using the Welch method~\cite{Welch} with public data around each event.

\begin{figure}[t]
 \includegraphics[width=\columnwidth]{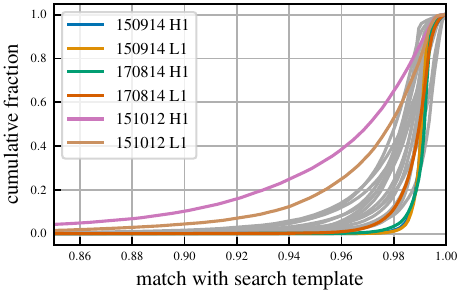}
 \caption{\label{fig:tmp_overlaps}
  The matches between the aligned-spin single template that we use in each of our searches for lensed
  counterparts of GWTC-1 events and the (precessing) public posterior samples~\cite{gwosc:GWTC-1} for the corresponding event.
  This is plotted individually for each event and for the \psd of each detector at
  the time of the event. We highlight GW151012, the observation with the lowest \snr for which
  the distributions of matches are broadest, and GW150914 and GW190814, the events with the highest
  \snr, for which the distributions are narrowest. All other events are plotted using gray lines.
 }
\end{figure}

The resulting matches are shown in Fig.~\ref{fig:tmp_overlaps}.
They are consistently high for the higher-\snr events
(e.g. a worst match of 95\% for GW150914)
and still quite high for the bulk of the posteriors for all other events.
For lower-\snr events, there is a tail
of a small number of the posterior samples for which the matches are lower,
in particular for GW151012 where the lowest match is 0.5. However,
even for GW151012 we find 90\% of samples with matches $>0.89$.
This indicates that a single aligned-spin template
matches well to the majority of each event's posterior
samples. Therefore our targeted single-template search provides the
ability to match well to most potential lensed counterpart signals
while greatly reducing the rate of background compared to a
full template bank search, as we demonstrate below.

\subsection{\label{sec:sig}Candidate identification and significance}

We use the \pycbc search pipeline~\cite{Usman:2015kfa,pycbc1.11.14}
to identify potential lensed counterpart images in the data.
Overall, our setup is quite similar to that of the original ``offline''
\pycbc analysis as reported in~\cite{LIGOScientific:2018mvr,DalCanton:2017ala}.
We briefly describe the search method here, highlighting differences
between our targeted search and the original \pycbc analyses.

First, matched filtering is performed separately for each LIGO detector,
and we record all single-detector candidates (``triggers'') above a minimum \snr of $4$.
This is a lower threshold than in the original GWTC-1 \pycbc search,
which used a threshold of 5.5.
This is particularly important for identifying candidates in times
when the sensitivities of the two detectors differ from each other.
For each single detector, candidate signal consistency tests are
applied \cite{PhysRevD.71.062001, Nitz:2017lco} and the single-detector statistic
described in \cite{Nitz:2017lco} is calculated.
For identifying these initial triggers we do not
use any information about the GWTC-1 events other than their apparent intrinsic
parameters---the observed component masses and spins.
We do however include the time-delay
between the observed candidate events and the original GWTC-1 observation
in an alternative approach to assessing
the significance of such candidates, as described below.
Other properties of the source, in particular the sky location,
only enter our method when performing our second step of Bayesian
hypothesis testing, as described in Sec.~\ref{sec:bayes}.

\begin{figure}[t]
 \includegraphics[width=\columnwidth]{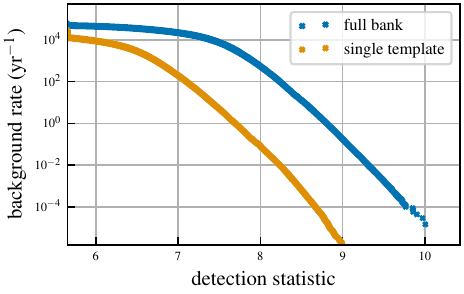}
 \caption{\label{fig:backgrounds}
  Background distribution,
  as a function of the detection statistic,
  for the single-template GW150914 counterpart search
  compared to a full template bank
  as used in the \pycbc search in~\cite{LIGOScientific:2018mvr}.
  The detection statistic is roughly proportional to the signal strain amplitude.
 }
\end{figure}

After using the search template to find a list of single-detector triggers,
we identify coincident candidates by testing for consistency in arrival time between detectors.
Each of these is assigned a ranking statistic (from \cite{Nitz_2017})
and we then compute their significance by comparing these values against a background
distribution.
This background is measured by performing a large number of unphysical time-shifts of the data sets~\cite{Allen:2005fk,Babak:2012zx}
and applying the same method as described above to find a new set of coincident triggers. The same ranking statistic can then be
calculated for this new set of background triggers.
Using this background we can estimate, for each candidate, the rate at which we expect to see events with
ranking statistic as loud or louder than that candidate to occur in the background;
we refer to this as the candidate's false-alarm rate.
With only a single search template, our targeted searches have much less freedom to
match background noise than the GWTC-1 searches, which used O(100,000) search
templates~\cite{LIGOScientific:2018mvr}. This means that a signal that does match
our single search template well will have a lower false-alarm rate than in the original
GWTC-1 search. This is illustrated in Fig.~\ref{fig:backgrounds} where we compare the mapping
from detection statistic to false-alarm rate when using only a single template against using
the full set of template waveforms as in the \pycbc GWTC-1 search~\cite{DalCanton:2017ala}.
For a background rate of 1 per year the required detection statistic is decreased by $\approx$1,
corresponding to $\sim$13\% smaller detectable amplitudes.

With the settings used in the GWTC-1 search,
\pycbc reports no more than one candidate signal within
a $\pm10$\,s window~\cite{Usman:2015kfa}.
This was motivated by the fact that the
event rate is small enough that it is highly unlikely
to observe two astrophysical signals from separate sources in a 10\,s window.
Using the same setting, we would not have been able to recover any lensed counterpart signals
with time delays shorter than 10\,s around each GWTC-1 event,
regardless of their strength.
To ensure that there are no lensed counterparts within a 10\,s window,
we perform additional reanalyses with this clustering criterion disabled
over these narrow time windows,
for all nine events besides GW170608.

In addition to the standard significance estimate of a candidate
in terms of a false-alarm rate from the time-shifted background,
it is also informative to include the candidate's time delay from the associated GWTC-1 event
when ranking possible lensed counterparts.
Astrophysical models for strong lensing favour shorter time delays,
but the specific distribution varies greatly between models
(see e.g.~\cite{Mao:1992}).
As a simple first step, we define a new ranking statistic
as the inverse false-alarm rate divided by the absolute value of the time delay. In the
case that the time delay is less than 10\,s, we divide by exactly 10:
\begin{equation}
\varrho = 
\begin{cases}
	\mathrm{IFAR} / |\Delta t|, & \text{if } |\Delta t| > 10 \,, \\
	\mathrm{IFAR} / 10, & \text{otherwise} \,,
\end{cases}
\end{equation}
where $\varrho$ is the new detection statistic, IFAR is the
inverse false-alarm rate and $\Delta t$ is the time-delay between the candidate
and the original GWTC-1 event.

This is equivalent to imposing a log-uniform prior on the time delay between the original event
and the lensed counterpart between 10\,s and the length of data searched and a uniform-in-time
prior between 0 and 10\,s time delays.
Such a prior does not directly follow from specific astrophysical expectations,
but is a simple choice to reflect our belief that an event with a time delay of minutes or hours is
more interesting than one with a time delay of months,
while not being too restrictive and setting the prior at
any time such that it would be effectively 0.

To produce a background for this new statistic we draw values of the inverse false-alarm rate
from our time-shifted background and combine them with a time-delay randomly drawn from the
analysed time. Using this background we can calculate a new delay-weighted false-alarm rate,
which is then converted into a \pval
\begin{equation}
p = 1 - \exp ( -T \cdot \text{FAR})
\end{equation}
where $p$ is the delay-weighted \pval;
$T$ is the the coincident time of the observation run used in the search;
and FAR is the false-alarm rate of the candidate.

This re-ranking could be repeated with any specific prior on time delays,
for example to test a particular lens and source population model with an
associated predicted time-delay distribution.
To facilitate such analyses we release our
full lists of coincident triggers~\cite{datarelease},
allowing others to perform such re-ranking accordingly.

It may also be informative to include information about the relative magnification compared
to the original GWTC-1 event. This could be achieved using a joint prior between time delay
and relative magnification. However, the time delay and magnification are not strongly correlated~\cite{Oguri:2010ns}
so we do not include such a prior in this first analysis.

It may be possible to reduce the background further by targeting the search towards signals
that would not be observable by the original search. However, we choose to leave this for
future work.

In our results, we therefore quote both
an inverse false-alarm rate obtained from the original ranking statistic
and the \pval obtained after applying the time-delay weighting.
Both the unweighted false-alarm rate and delay-weighted \pval
are computed independently for each search,
and do not include an overall trials factor.
Hence, under the null hypothesis of no additional \gw signals in the data
we typically expect one candidate
with a \pval of $\lesssim 0.1$ from our combined set of 10 searches.

\subsection{\label{ssec:search_sensitivity}Search sensitivity}

To assess the sensitivity of our search to lensed counterpart images we use simulated
signal waveforms added to the detector data. For each of our 10 searches we add a set
of simulated signals whose parameters are drawn from the full GWTC-1 posterior samples~\cite{gwosc:GWTC-1}
and generated using the precessing IMRPhenomPv2 waveform model.
For each of the 10 injection sets the coalescence times of the simulated signals are varied
uniformly covering a full month of data either side of the original GWTC-1 event.
The amplitude of each signal is also multiplied by a scale factor, equal to the square root
of the lensing magnification (relative to the original GWTC-1 event).
The scale factor is drawn from a uniform distribution between 0.1 and 10 in order to cover
the range of signals recoverable by the search, taking into account changes in detector sensitivity
over time.

The sensitivity of \gw transient searches is often evaluated in terms
of sensitive distance, sensitive volume or sensitive volume multiplied by accumulated time.
However, it is not straightforward to discuss these measures
when dealing with lensed signals, since there will
be a large uncertainty on the true luminosity distance to the source.
We can quote such a standard sensitivity estimate however,
if we consider our simulated signals with scaled amplitudes not as magnified through
lensing, but as having different luminosity distances as for unlensed signals.
We can then quote the ``sensitive distances'' of each search,
as listed in Table~\ref{table:searches}.
These sensitive distances are calculated by applying a detection threshold on the
false-alarm rate of 1 per year and finding the detection efficiency in
a number of distance bins. The volume of each distance bin is multiplied by the search efficiency
before being summed and converted to a sensitive distance. We do not use the delay-weighted
\pval[s] when computing these efficiencies.

\begin{figure}[t]
 \includegraphics[width=\columnwidth]{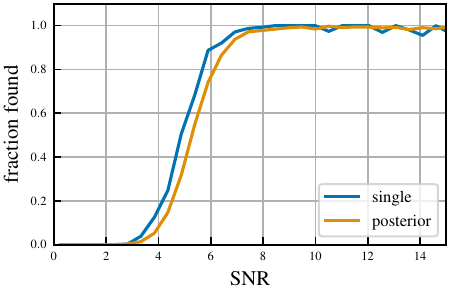}
 \caption{\label{fig:inj_single_pos_GW151012}
  Fraction of recovered signal injections for GW151012
  for two injection sets:
  one ('single') with all parameters fixed to those of the search template,
  and one ('posterior') with parameters drawn from the full GWTC-1 posterior.
  An injection is counted as recovered if found with an inverse false-alarm rate greater than one year.
  The horizontal axis gives the injection \snr
  defined as the smallest of the two optimal \snr[s] in each detector.
 }
\end{figure}

To compare the sensitivity of our searches, where we use only a single template waveform,
to that of the idealised case in which the template bank has 100\% match with the full posterior,
we create a second set of injections with parameters identical to the \map values
for each search, still searching for them with the same fixed single template. In this case
the search template will always match well with the simulated signal.
As an example, we consider GW151012, the lowest-\snr event.
Fig.~\ref{fig:inj_single_pos_GW151012} shows the difference in recovery
for the two injection sets,
corresponding to a loss of sensitive distance of $\sim$12\% compared to the idealised case.
This is the largest difference between the two injection sets for all 10 events,
with losses in sensitive distance of $\sim$3 -- 9\% for the others.
This again demonstrates that searching with a single template provides good coverage of the events' posteriors,
with the majority of events having a loss of $\lesssim$ 7\% compared to the idealised case.
The sensitive distances for both injection sets for all 10 searches are listed in Table~\ref{table:searches}.

This measure provides an estimate of the loss of sensitivity
due to the use of a single template to cover the posterior of each event.
However, in practice a search using more templates
to cover the full posterior will also increase the rate of background triggers, reducing the sensitivity.
Additionally, in this test the injected signals are a perfect match to the search template,
whereas a practical template bank will still have some mismatch
due to the discrete placement of templates in any bank,
which must be balanced with the number of templates required to cover the posterior.
There will also be some mismatch between the true signal and the employed waveform
models due to the lack of precession in the current search,
as well as a small effect due to the limited accuracy of numerical-relativity calibration.
An optimal template bank construction along these lines will be an interesting avenue for future research,
but, as demonstrated, our extreme single-template choice can already produce very sensitive results.

\subsection{\label{sec:bayes}Lensed events or separate gravitational-wave signals? }

If any of the single-template searches recover significant candidates,
we will have to ask the question whether these are actually the lensed images we are looking for.
In other words, if we assume that any given candidate event is astrophysical,
what is the probability that it is a lensed counterpart of the original event,
compared to the probability that it is an independent \gw signal?

In \cite{Hannuksela:2019kle}, the same question for pairs of GWTC-1 events was
answered through \kde-based overlap integrals of posterior distributions
from single-event Bayesian inference,
following the method introduced by \cite{Haris:2018vmn}.
This approach is somewhat limited by the practical problems
in robustly constructing high-dimensional \kde[s],
and hence not enforcing the full constraints on the consistency of the two events
as expected under the lensing hypothesis.
Instead, here we perform this test through joint Bayesian inference on the pairs of events,
fully coherently combining the data from all available detectors for the two relevant times.
To compare the lensed and unlensed hypotheses
we calculate the evidences $\mathcal{Z}$ for each and use these to find the Bayes' factor (evidence ratio)
$\blu$ between the two.

A similar approach of ``joint parameter estimation'' on candidate pairs of lensed events
was previously suggested in the context of space-based \gw observations by~\cite{Seto:2003iw}
and for the case of LIGO-Virgo observations similar methods have been developed independently~\cite{UWMPREP, RicoPREP},
based on the LALInference and bilby parameter estimation packages~\cite{Veitch:2014wba,Ashton:2018jfp}.

In the unlensed hypothesis the two candidates $i = 1,2$ can be described with independent sets of parameters
$\theta_{i} = \{m_1, m_2, a_{1z}, a_{2z}, d, \iota, \alpha, \delta, \psi, \phi_c, t_c\}$,
where $\alpha$ and $\delta$ are the right ascension and declination of the source,
$\iota$ is the inclination, $\psi$ is the polarization angle,
$\phi_c$ is the coalescence phase, and $t_c$ is the time of coalescence.
The evidence can therefore be calculated as the product of the evidences
for each event $\ZU = \mathcal{Z}_{1} \cdot \mathcal{Z}_{2}$, where each
\begin{equation}
\mathcal{Z}_{i} = \int P(d_{i} | \theta_{i}, \mathcal{I}) P(\theta_{i} | \mathcal{I}) d\theta_{i}
\end{equation}
and $d_{i}$ is the dataset for event $i$; $\mathcal{I}$ represents any implicit assumptions made within
the model, e.g. that the data is Gaussian and locally stationary; $P(d_{i} | \theta_{i}, \mathcal{I})$ is
the likelihood of producing the data given the parameters; and $P(\theta_{i} | \mathcal{I})$ is
the prior on the parameter space.

To calculate the evidences we perform nested sampling
using \pycbcinf~\cite{Biwer:2018osg,pycbc1.13.8}
with the \dynesty~\cite{2020MNRAS.493.3132S} sampler
and the aligned-spin IMRPhenomD waveform model~\cite{Husa:2015iqa,Khan:2015jqa},
again not considering mis-aligned spins as there is no evidence of precession for any of the GWTC-1 events.
We use a fixed prior for all candidates:
$t_c$ uniform within 0.2\,s around the time reported by the search,
component masses uniform within $[5,120]\,\Msun$ (detector frame),
spin magnitudes uniformly distributed in $[0,0.99]$, with an equal probability
of being aligned or anti-aligned with the orbital angular momentum,
distance uniform in $[10,5000]$\,Mpc, and $\psi$ uniform in $[0, 2\pi]$.
The pairs of $(\iota,\phi_c)$ and $(\alpha,\delta)$
are both chosen so that they are uniformly distributed on a sphere.
The \psd is calculated using 1024\,s of data centred on the time reported by the search.

Under the lensed model the two signals share the same astrophysical origin
and therefore share the same intrinsic parameters. The typical change in sky position
between lensed signals is far smaller than the resolution of \gw detectors
and $(\alpha,\delta)$ can therefore also be treated as shared parameters.
The two candidates therefore have a set of common parameters
$\theta^{\prime} = \{m_1, m_2, a_{1z}, a_{2z}, d, \iota, \alpha, \delta, \psi\}$
with only the coalescence phase, time of coalescence and
relative magnification changing between events:
$\theta_{i} = \{\phi_c, t_{c}, \murel\}$,
where $\mu_{\mathrm{rel}} = \mu_{1}/\mu_{i}$
is the magnification of the additional candidate
relative to the original GWTC-1 event.
The evidence is therefore given by
\begin{equation}
\label{eq:ZL}
\begin{split}
\ZL = \int & P( d_{1} | \theta^{\prime}, \theta_{1}, \mathcal{I} ) \;
	P( d_{2} | \theta^{\prime}, \theta_{2}, \mathcal{I} ) \\
	& \cdot P( \theta^{\prime}, \theta_{1}, \theta_{2} | \mathcal{I} ) \;
	d\theta^{\prime} d\theta_{1} d\theta_{2} \,.
\end{split}
\end{equation}

This lensed evidence is also calculated using nested sampling.
For each sample, we draw a set of common parameters $\theta^{\prime}$ and a set of independent 
parameters $\theta_{i}$ for each candidate.
The shared parameters use the same priors as in the unlensed case.
In addition, the first event has a uniform prior on $\phi_c$ in $[0, 2\pi]$,
while the coalescence phase of the second signal,
in the case of an aligned-spin system undergoing circular inspiral,
can be shifted relative to the first \cite{Dai:2017huk}
by values of $\{0, \pi/4, \pi/2, -\pi/4\}$;
these values are each given an equal prior weight.
The prior on $t_c$ for the first event is uniform in a 0.2\,s window around the original time reported in GWTC-1,
while the prior on $t_c$ for the second candidate is uniform in a 0.2\,s
around the trigger time reported by our search.
Finally, the relative magnification is 1, by definition, for the original GWTC-1 event,
and uses a uniform prior on $\sqrt{\murel}$ between $[0.1, 10]$ for the second candidate.

We then combine the shared parameters and independent parameters for each signal
and use these to produce a separate waveform for each.
These waveforms are then used to evaluate the two likelihoods in Eq.~\eqref{eq:ZL},
which we then combine to calculate the total likelihood for the lensed model.
This method allows us to calculate the lensed evidence directly using nested sampling.
Finally we combine the two evidences to calculate the Bayes' factor $\blu = \ZL/\ZU$.

These Bayes' factors are only meaningful if both events are of an astrophysical origin.
In order to produce a Bayes' factor comparing the lensed hypothesis to the unlensed hypothesis
without assuming that the candidates are astrophysical, a more detailed model for terrestrial
noise sources would be required.

The Bayes' factors produced by this analysis do not only test if two candidates are consistent
with a single set of parameters, but also implicitly test if two independent astrophysical events are
likely to occur with these parameters given the priors.
A pair of events with parameters consistent with one
another will produce a higher Bayes' factor when occurring in a region of low prior support than in a region
of high prior support,
matching the expectation that multiple independent astrophysical events should
be observed more frequently in regions of high prior support.
For this reason the Bayes' factors produced by this analysis will be highly prior-dependent,
with a wider prior favouring the lensed hypothesis more.

In the future it will be possible to use the large
number of \bbh[s] observed in O3 to produce an astrophysical prior for this analysis,
but for the current analysis
we will see in the next section that the standard wide uniform priors are already sufficient
to reject the possible candidate pairs that we find in this case.

\section{\label{sec:candidates}Recovered candidates}

\squeezetable
\begin{table*}[t]
\setlength{\tabcolsep}{4pt}
\begin{tabular}{lS[table-format=1.7]c@{\hskip 2pt}p{4pt}cc@{\hskip 2pt}p{4pt}S[table-format=3.6]r}
\toprule
delay-weighted $p$-value & {false-alarm rate$^{{-1}}$ [yr]} & UTC time & & known event? & found by & & {$|\Delta t|$ [d]} & $\log_e(\blu)$
\\[0.5\baselineskip]
\midrule \midrule
0.16 & 0.166     & 2017-07-30 08:05:26.8 &   &          & GW170729 &   & 0.548     & -0.77 $\pm$ 0.40   \\
0.29 & 0.0000913 & 2017-01-04 10:12:57.9 &   &          & GW170104 &   & 0.000687  &  4.43 $\pm$ 0.33   \\
0.37 & 0.497     & 2017-08-04 14:57:29.3 & * &          & GW170809 & * & 4.73      & -2.25 $\pm$ 0.32   \\
0.4  & 0.000550  & 2017-07-29 19:05:05.9 &   &          & GW170729 &   & 0.00598   & -1.43 $\pm$ 0.26   \\
0.46 & 0.000465  & 2015-09-14 10:04:34.7 &   &          & GW150914 &   & 0.00960   &  4.61 $\pm$ 0.37   \\
0.48 & 0.000241  & 2017-07-29 18:51:31.4 &   &          & GW170729 &   & 0.00345   & -2.74 $\pm$ 0.18   \\
0.48 & 0.000131  & 2017-01-04 10:14:56.3 &   &          & GW170104 &   & 0.00206   &  3.16 $\pm$ 0.25   \\
0.86 & 2.53      & 2017-04-01 08:19:53.0 &   &          & GW170809 & * & 130       &  3.41 $\pm$ 0.59   \\

\midrule \midrule
0.0067 & 2246    & 2017-01-21 21:25:36.6 &  & GW170121 & GW170818 & * & 208       & -2.55 $\pm$ 0.92   \\
0.064 & 191      & 2017-01-21 21:25:36.6 &  & GW170121 & GW170823 &   & 214       & -8.05 $\pm$ 0.52   \\
0.097 & 1.57     & 2017-07-27 01:04:30.0 &  & GW170727 & GW170729 &   & 2.74      & -1.98 $\pm$ 0.48   \\
0.37 & 15.5      & 2017-03-04 16:37:53.4 &  & GW170304 & GW170729 &   & 147       & -2.41 $\pm$ 0.40   \\
0.48 & 12.4      & 2017-03-04 16:37:53.4 &  & GW170304 & GW170823 &   & 172       & -7.30 $\pm$ 0.55   \\
0.99 & 1.14      & 2017-03-04 16:37:53.4 &  & GW170304 & GW170818 & * & 166       & -5.54 $\pm$ 0.76   \\

\bottomrule
\end{tabular}
 \caption{\label{table:candidates}
  The most significant recovered candidates from our 10 single-template searches,
  sorted by their
  delay-weighted \pval
  which prefers short time delays
  (first column),
  cut at $\leq0.5$.
  Candidates with an inverse false-alarm rate
  (second column, derived from the original ranking statistic)
  $\geq 1$ year
  are also included,
  regardless of their time delay.
  The third column gives the candidate end time (UTC)
  and the fourth column gives the event name for those candidates
  already published by~\cite{Venumadhav:2019lyq,Nitz:2019hdf}
  (listed separately in the lower half of the table).
  The GWTC-1 event whose counterpart search found the candidate
  is listed in the fifth column,
  and the absolute value of the time delay
  is given in the sixth column.
  Candidates that are themselves listed in GWTC-1 are excluded;
  see Table~\ref{table:candidates-all} in the \supplement
  for those.
  The final column is the log Bayes' factor (evidence ratio)
  comparing the
  lensed vs unlensed hypothesis, assuming both events are
  astrophysical and using nested sampling to
  compute the evidence when the two events have a shared set of
  parameters and when their parameters are independent
  (see Sec.~\ref{sec:bayes}). Times and events that have Virgo data
  available have been marked with an asterisk (*); this data has
  been used only when calculating the log Bayes' factors.
  }
\end{table*}

In this section we will show the results of our 10 single-template searches,
along with the results of Bayesian inference carried out for each of the
resulting candidates.

Several of our searches recover other GWTC-1 events
with high significance;
details of these cross-matches are provided in Table~\ref{table:candidates-all} in the \supplement.
This is an expected result of the well-known clustering of
high-mass detections in a small part of the overall parameter
space~\cite{LIGOScientific:2018jsj,Venumadhav:2019lyq,Nitz:2019hdf}.
As already demonstrated by \cite{Hannuksela:2019kle},
none of these pairs of GWTC-1 events are viable candidates for strongly-lensed double images
after considering their relative time delays and
the degree of posterior overlap over all relevant parameters.

\begin{figure*}[th]
 \includegraphics[width=\textwidth]{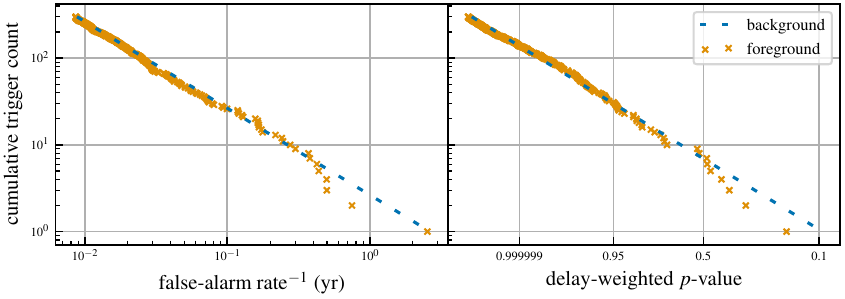}
 \caption{\label{fig:null}
  Combined results from our 10 single-template searches.
  Left panel: The cumulative count of coincident triggers with inverse false-alarm rates
  less than or equal to a given value.
  Right panel: The cumulative count of coincident triggers with delay-weighted \pval[s]
  greater than or equal to a given value.
  The dashed line is the distribution of background triggers produced from
  time-shifted detector data with randomly drawn time delays.
  The crosses are foreground triggers, excluding those from GWTC-1
  and those reported in \cite{Venumadhav:2019lyq}.
  This set of foreground triggers is fully consistent with the null hypothesis.
 }
\end{figure*}

Table \ref{table:candidates} lists the most significant candidate events from
our search after removing pairs of GWTC-1 events.
We separate this table into
new candidates
and those already reported by~\cite{Venumadhav:2019lyq} and later by~\cite{Nitz:2019hdf}.

The most significant new candidate by inverse false alarm rate has a value of 2.53 years,
this is consistent with the expectation that we would observe one noise event with
a similar significance, as the 10 searches covered a total of $\sim$2.6 years of data.
Additionally, the most significant new candidate when ordered by \pval has a value of
0.16, due to no trial factor being included in the statistic one event with a \pval
of $\lesssim 0.1$ is expected from the combined 10 searches.
The set of new candidates is therefore consistent with the null hypothesis
when ranked either by inverse false-alarm rate or \pval. This can also be seen in
Fig.~\ref{fig:null}, showing that the rate of new foreground events is consistent with
the empirically measured background.

However, we do recover three of the events from~\cite{Venumadhav:2019lyq}
(GW170121, GW170304, GW170727)
with false-alarm rates of less than 1 per year,
some of them from multiple searches.
These were not considered in the test for lensed pairs in~\cite{Hannuksela:2019kle}.
For GW170121 and GW170304, the large time delays from their matching GWTC-1 events
(5 to 7 months)
already suggest that these pairs are unlikely due to lensing
(at least by the more common galaxy lenses),
but more probably come from two unrelated sources with similar characteristics.
On the other hand,
the time delay between GW170727 and GW170729 is relatively short,
warranting further investigation.

As discussed in Sec.~\ref{sec:bayes} we perform Bayesian inference
on all candidate pairs in Table~\ref{table:candidates} in order to
calculate Bayes' factors $\blu$ comparing
the hypothesis that they are lensed counterparts
against the hypothesis that the two events come from different sources.
Positive values indicate support for
the lensed hypothesis. The analysis is performed multiple times in order
to estimate the error on the Bayes' factor and
the results of this analysis are included in Table~\ref{table:candidates}.

The $\log_e\blu$ values for four of the new low-significance candidates in the first half of the table
are nominally supportive of the lensed hypothesis.
However, these must be considered in the context of several mitigating factors.
First, the expected lensing rate at O1/O2 sensitivities is very small
under standard astrophysical assumptions~\cite{Ng:2017yiu,Oguri:2018muv},
meaning that very large Bayes' factors, $\mathcal{O}(1000)$,
would be needed to obtain high posterior odds
after factoring this rate in as an explicit prior between the two hypotheses.
Second, these results are not independent of the initial search stage. By first testing
for consistency in the intrinsic parameters, selection effects are introduced that will
favor larger values of $\log_e\blu$.
As discussed in Sec.~\ref{sec:bayes} the wide priors on masses used in this analysis will also favor the
lensed hypothesis if the astrophysical distribution
on masses~\cite{LIGOScientific:2018jsj} is much narrower than this.
Most importantly, we have already shown that the set of candidates in the first
half of the table are fully consistent with the background noise model,
while the Bayes' factors calculated are only meaningful
if both candidates are real astrophysical events.
Our conclusion, therefore, is that these are most likely noise events,
and the positive $\log_e\blu$ values are spurious.

For the events in the second half of the table,
which have been previously discovered~\cite{Venumadhav:2019lyq,Nitz:2019hdf}
as independent \bbh candidates,
all $\blu$ values favor the unlensed hypothesis.
We therefore find no evidence for strong lensing present
between the \bbh[s] reported in GWTC-1
and \cite{Venumadhav:2019lyq,Nitz:2019hdf}.
This analysis therefore extends the results from~\cite{Hannuksela:2019kle}
to demonstrate that no pairs of candidates from O1 and O2
have evidence supporting the presence of strongly lensed double images.

While negative, these Bayesian inference results are still useful
in order to investigate the types of candidates that
subthreshold searches for lensed counterparts return.
In future searches of this type,
it will be a continuing challenge to confidently identify a lensed
counterpart observation as opposed to observing a pair of similar but separate GW events.
However, the $\blu$ can at least be used to rule out pairs that strongly disfavor
the lensed hypothesis.
More detailed analyses taking into account detailed lens modelling~\cite{10.1093/mnras/staa2577}
or even targeted follow-up campaigns~\cite{Smith:2018kbc}
may then become feasible on any strong remaining candidates.
Also, as our knowledge of the astrophysical distribution of GW source
systems improves, we will be able to better constrain the priors that go into this analysis
and hence better distinguish promising candidates for lensed pairs
from chance coincidences in parameter space.

\section{\label{sec:uls}Implications of the absence of clear counterpart candidates}

We now explore the astrophysical implications of
the null result of our searches, based on their estimated sensitivity.
To find the expected recovery rate for lensed images in our searches we use simulated
signals added to the \gw strain data before matched filtering (``injections'').
Of the injection sets described in Sec.~\ref{sec:templates},
in the following we always use the injections drawn from the full
precessing posterior of the original GWTC-1 events~\cite{gwosc:GWTC-1}.
Their strain amplitudes are multiplied by a scale factor
\mbox{$\sqrt{\murel}=\sqrt{\mu_0/\mu_\mathrm{inj}}$},
where $\mu_0$ and $\mu_{\mathrm{inj}}$ are
the absolute magnifications of the primary and injected signals respectively.

We combine two injection sets, one uniformly covering a full month of data on either
side of the original GWTC-1 event with \mbox{$\sqrt{\murel}\in[0.1,10]$}, and one focussing
on a smaller window 2 hours either side of the event with \mbox{$\sqrt{\murel}\in[1/3,3]$}
in order to more accurately calculate recovery rates for short time delays.
Any injection made outside of the analysed times for either of the two detectors
is considered to be missed to account for the detector duty cycle.

\begin{figure*}[th]
 \includegraphics[width=\textwidth]{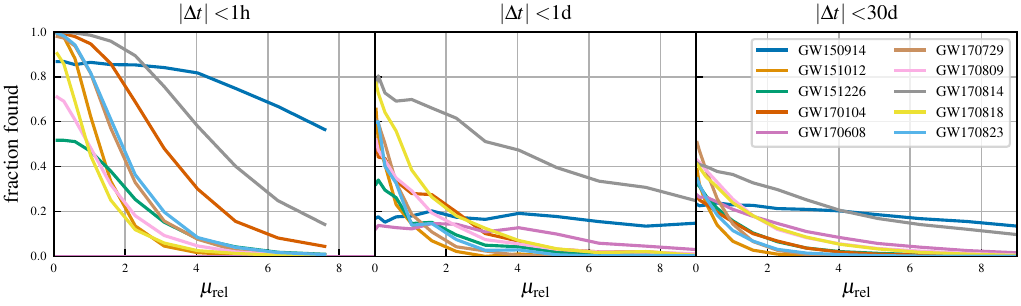}
 \caption{\label{fig:pdet_from_injs_multi_event}
  The probability of finding lensed images of the GWTC-1 events
  in each of the 10 corresponding searches
  as a function of relative magnification
  \mbox{$\murel=\mu_{0}/\mu_\mathrm{inj}$},
  for several ranges of time delays $\delay$.
  This is calculated using simulated signals drawn from each event's posterior
  with thresholds of an inverse false-alarm rate $\geq1$\,year or
  delay-weighted $p$-value $<0.16$.
 }
\end{figure*}

The recovery rate of these injections
yields an estimate for the probability
of finding a lensed image as a function of
magnification ratio $\murel$ and time delay $\delay$.
These results are summarized in Fig.~\ref{fig:pdet_from_injs_multi_event},
where we have now used a threshold based on the actual search results:
to be counted as recovered, an injection must be
more significant than anything in the top half of Table~\ref{table:candidates},
i.e. have an inverse false-alarm rate $\geq1$\,year
or a delay-weighted $p$-value $<0.16$.
Along the $\murel$ dimension,
we find that
recovery mostly depends on the strength of the original \gw event.
Along the time dimension, recovery is mostly limited
by the $\lesssim50$\,\% coincident livetime of the LIGO detectors during O1 and O2.

Any astrophysical interpretation of the absence of convincing counterpart candidates
depends on the choice of priors for the true (unlensed) high-redshift \bbh source population
and the properties of lenses in the Universe.
We provide supplementary data~\cite{datarelease} for the sensitivity of all 10 searches
so that other authors may explore the implications of our search results
under different choices for these priors.

\subsection{\label{sec:extreme}Extreme lensing hypothesis}

Here, we demonstrate one application of these results,
testing a strict interpretation of the idea from~\cite{Broadhurst:2018saj}:
What if \emph{all} of the high-mass GWTC-1 \bbh events
really came from lighter objects
at higher redshifts?
To be more specific, we phrase the test hypothesis as:
\textit{The intrinsic component masses of any \bbh in the Universe cannot be larger
than $15\,\Msun$.
All apparently heavier \gw events are due to lensing}.

Let us start from the following quantities under
the standard no-lensing hypothesis ($\unlens$)
from GWTC-1~\cite{LIGOScientific:2018mvr}:
primary source-frame masses $m_{1,\unlens}$,
redshifts $z_\unlens$
and luminosity distances $d_\unlens$.
We then define the extreme lensing hypothesis ($\lens$) as
there being no merging \bbh[s] in the Universe
with component masses over 15$\Msun$;
we implement this by fixing the intrinsic (source-frame) primary mass of each event to
\mbox{$m_{1,\lens} = 15\,\Msun$}.\footnote{In \cite{Broadhurst:2018saj},
the $15\,\Msun$ limit was originally phrased in terms of chirp mass.
But since it is motivated by the observed masses for galactic black holes
with a reference to~\cite{Dominik:2013tma},
which in turn refers to the \emph{individual} black hole masses
in Cyg~X-1 and GRS~1915,
we reinterpret the hypothesis in terms of component masses.}
Having all primary masses equal
to exactly $15\,\Msun$ is an unlikely source distribution, but including a distribution of lower masses
would only increase the magnifications required to produce the same detector frame masses.
As discussed below, this would decrease the probability of no additional lensed images being
present, therefore increasing the chance of observing these counterparts
and strengthening the arguments below.

We can then easily obtain the `corrected' redshifts
under the lensing hypothesis as
\begin{equation}
 z_\lens = \frac{m_{1,\unlens}}{15\,\Msun} (1+z_\unlens) - 1 \,.
\end{equation}
These are then converted to luminosity distances
under the standard Planck cosmology~\cite{Ade:2015xua},
and the corresponding lensing magnification factors are
\mbox{$\mu = (d_\lens/d_\unlens)^2$}.
Table~\ref{table:lens-hypo-params} lists the results
for the eight heaviest \bbh sources
(with median \mbox{$m_1>15\,\Msun$}),
using median results from~\cite{LIGOScientific:2018mvr} for simplicity.

GW151226 and GW170608 have median \mbox{$m_1<15\,\Msun$}
and hence are already consistent with the hypothesis
that all black holes have masses below $15\,\Msun$,
so we do not include these in our test.
GW151012 at a median \mbox{$m_1\approx23\,\Msun$} is included in the table for completeness
but also not used in the test,
because its required magnification of 30 does not fall
into the sufficiently extreme regime for the approximations discussed below,
which require magnifications greater than $\sim$100.
As we will see, the result of the test is also already sufficiently strong
from just considering the remaining seven higher-mass signals.

For the seven remaining signals the ratio of intrinsic and observed masses yields
corrected redshifts and luminosity distances,
from which the required magnifications are between 100 and 800
with GW150914 having the largest magnification.
(Again, see Table~\ref{table:lens-hypo-params}.)

\begin{table*}[t]
\begin{tabular}{cccccccccc}
\toprule
event    & $m_{1,\unlens}$ [$\Msun$] & $z_\unlens$ & $z_\lens$ & $d_\unlens$ [Gpc] & $d_\lens$ [Gpc] & $\mu$ & $P_\mathrm{single}$  & $P_\mathrm{missed}$ & $P_\mathrm{found}$\\
\midrule
GW150914 & 36                        & 0.09        & 1.6       & 0.4               & 12              & 800  & 0.020 & 0.031 & 0.950 \\
GW151012 & 23                        & 0.21        & 0.9       & 1.1               &  5.8            &  30  & ---   & ---   & ---   \\
GW170104 & 31                        & 0.20        & 1.5       & 1.0               & 11              & 130  & 0.046 & 0.127 & 0.832 \\
GW170729 & 50                        & 0.49        & 4.0       & 2.8               & 37              & 170  & 0.040 & 0.116 & 0.848 \\
GW170809 & 35                        & 0.20        & 1.8       & 1.0               & 14              & 200  & 0.038 & 0.072 & 0.893  \\
GW170814 & 31                        & 0.12        & 1.3       & 0.6               &  9.3            & 260  & 0.034 & 0.043 & 0.925 \\
GW170818 & 35                        & 0.21        & 1.8       & 1.1               & 19              & 200  & 0.038 & 0.051 & 0.913  \\
GW170823 & 40                        & 0.35        & 2.5       & 1.9               & 21              & 130  & 0.046 & 0.127 & 0.833 \\
\bottomrule
\end{tabular}
 \caption{\label{table:lens-hypo-params}
  Parameters of the eight heaviest GWTC-1 events,
  reinterpreted under the extreme lensing hypothesis ($\lens$)
  that they all should have intrinsic primary masses of
  \mbox{$m_{1,\unlens} = 15\,\Msun$}.
  Unlensed parameters ($\unlens$) correspond to the median values from~\cite{LIGOScientific:2018mvr}.
  Luminosity distances are obtained under a standard Planck cosmology~\cite{Ade:2015xua}.
  $P_\mathrm{single}$ is the probability of an event of this magnification
  not having a comparably bright counter image (with \mbox{$1/3\leq\murel\leq3$}).
  $P_\mathrm{missed}$ is the probability that our search fails
  to recover the counter image in the LIGO data
  (with delay-weighted $p$-value below 0.16).
  $P_\mathrm{found}$ is the probability of our search recovering a counter image
  under the extreme lensing hypothesis.
  GW151012 is excluded from the analysis as we only consider systems with magnification $>100$.
  }
\end{table*}

\subsection{\label{sec:astro}Astrophysical priors on magnification and time delay}

Lensing events with magnifications greater than 100 are rare in the Universe,
with the distribution of magnifications following \mbox{$P(\mu) \sim \mu^{-3}$},
due to the small area in the lens plane
where high $\mu$ can be produced.
Still, such values are possible, particularly for point sources.
For example a star at redshift 1.5 has been observed
with a magnification of $\sim$2000 \cite{kelly2018}.
The catastrophe theory of strong gravitational lensing~\cite{Petters:2001stgl}
shows that images with very high magnification are formed when the source lies either:
(i) just inside a fold catastrophe,
forming a pair of images with the same brightness;
(ii) just inside a cusp catastrophe,
forming a triplet with one image twice as bright as the other two;
(iii) or just outside the cusp catastrophe,
forming a single highly magnified image \cite{Schneider:1992}.
Higher order catastrophes can produce more complicated configurations \cite{keetonmaowitt2000,evanswitt2001}
but are extremely rare \cite{orbandexivry2009,collettbacon16}.
The time delays between the multiple highly magnified images are extremely short,
typically seconds to at most hours,
with \mbox{$\Delta t \sim \mu^{-3}$}.

The lensing mass of galaxies is well approximated by singular isothermal ellipsoids \cite{auger2010}.
In this case the fraction of highly magnified images without a comparably bright counter-image
is given by \cite{kormann1994}
\begin{equation}
P_\mathrm{single} = \frac{1}{1+ 4 \pi \frac{15}{16 \sqrt{6}}\frac{\sqrt{1-q^2}}{1+q}\mu^{1/2}}
\label{eq:ratio}
\end{equation}
where $q$ is the axis ratio of the lens.
Thus, unless the lens is very close to spherical ($q=1$),
highly magnified images are unlikely to occur without a bright counter-image.
In the case of a lensed \bbh,
we do not know the specific lens and cannot measure $q$ directly.
Instead we must marginalize over the population of all potential lenses in the Universe.
We use the lens population model of \cite{collett2015}
to realize the population of gravitational lenses in the Universe.
This model assumes lens galaxies are singular isothermal ellipsoids.
The lenses are randomly distributed with uniform comoving number density out to redshift 2. Lens masses are derived from the SDSS velocity dispersion function \cite{choiparkvogeley}. 
Lens ellipticities are drawn from the velocity dispersion dependent probability density function fit to SDSS \cite{collett2015}.
To perform the marginalisation, we draw $10^8$ lenses from the \cite{collett2015} population and weight them by their strong lensing cross section (proportional to the Einstein radius squared) for the source plane of each \bbh.
For each lens, we calculate the probability of seeing a single bright image using Eq. \eqref{eq:ratio}.
The final probability of seeing a single image is thus
\begin{equation}
P_\mathrm{single} = \sum_{i} \Theta_{E,i}^2 P_{\mathrm{single},i} / \sum_{i} \Theta_{E,i}^2,
\label{eq:maginalisedsingleimage}
\end{equation}
where $\Theta_{E,i}$ is the Einstein radius of each lens.

Using Eq.~\eqref{eq:maginalisedsingleimage} to marginalize over the lens population
we find that only 2\% of images with \mbox{$\mu>800$},
or 4\% with \mbox{$\mu>100$},
have no counter-image with comparable magnfication (\mbox{$1/3\leq\murel\leq3$}).
We define the probability of not having such a bright counter image
as $P_\mathrm{single}$ in Table~\ref{table:lens-hypo-params}.

The same lens population model allows us to numerically infer
the time delay and magnification ratio between counter-images.
For each putative lensed BBH event,
we realise $10^5$ lens systems weighted by their lensing cross section
given the true source redshift.

For each lens we draw a random position from a uniform distribution on the image plane.
For each position we infer the magnification of a strongly lensed image forming at that location.
We do this repeatedly until we find 1000 image positions for each lens
that have a magnification within 10\% of the putative magnification.
To find the unlensed source positions that produce these highly magnified images, we trace the highly magnified images back onto the source plane using the lens equation\footnote{The lens equation relates the observed and true angular positions of a lensed source.
The difference between these two positions is the reduced deflection angle, which is sensitive to the observed image position, the mass distribution of the lens and to the angular diameter distances between observer, lens and source.}.
This gives a set of source positions for each lens that produce an image of the correct magnification.
For each source position, we use the lens equation to find the locations on the image plane where the counter images form. For each counter image we calculate the magnification and the time delays relative to the image with the putative magnification.
This procedure gives a list of $10^8$ lensed \bbh events with time delays and magnification ratios relative to the highly magnified image. Assuming the lensing hypothesis, each of the events in this list could have produced the observed highly magnified event.

This model shows that for the highly magnified images
as required by the extreme lensing hypothesis,
90\% (99\%) of the comparably bright counter-images occur within e.g.
5 (45) minutes for GW170823
and 2 (15) seconds for GW150914.

\subsection{\label{sec:pfound}Results of the hypothesis test}

For each simulated highly magnified lensing event,
the probability of missing the counter-image is given by one minus the recovery fraction
(calculated using the same thresholds as for Fig.~\ref{fig:pdet_from_injs_multi_event},
already including the duty factors of the detectors during O1 and O2)
for injections with the correct time delay and strain ratio relative to the observed BBH event.
Marginalizing over all of the simulated image pairs gives
the overall probability of missing a comparably bright counter-image, $P_\mathrm{missed}$.
The probability of seeing a counter-image is thus 
\begin{eqnarray}
P_\mathrm{found} &=& 1 - P_\mathrm{single} -(1-P_\mathrm{single})P_\mathrm{missed} \nonumber \\
&\approx& 1- P_\mathrm{single} - P_\mathrm{missed} \,.
\end{eqnarray}
This yields recovery fractions ranging from 95\% for GW150914 to 83\% for GW170104.
These values are shown for each event in Table~\ref{table:lens-hypo-params}.

The product of $1-P_\mathrm{found}$ for the seven heaviest events is $1 \times 10^{-7}$.
This is the probability of missing the counter-images for all of the 7 events
assuming they are all lensed by the magnification corresponding to a maximum component mass of 15 $\Msun$.
If the unlensed masses were lower,
the required magnifications increase,
decreasing both $P_\mathrm{single}$ and $P_\mathrm{missed}$,
thus in turn driving $P_\mathrm{found}$ further towards 1.

The model presented here does not include lensing by clusters.
The mean Einstein radius in the model is 0.7 arcseconds,
and the time delays are proportional to the Einstein radius,
so even if clusters dominate the lensing cross section,
the expected time delays would only increase by a factor of a few
as the Einstein radius of typical clusters is $\sim$5 arcseconds \cite{zitrin2012}.

The model also does not account for deviations from isothermality.
This introduces a small change in the constant of proportionality in the time delays
between highly magnified image pairs.
To assess the potential size of these systematics, we rerun our pipeline,
but assuming all time delays are 10 times longer than in the fiducial model:
in this scenario the probability of missing all of the counter-images increases to $5 \times 10^{-7}$.
Therefore, possible systematics introduced by cluster lenses and deviations from isothermality will not
change the result significantly.
Thus, the observed lack of detecting any lensed counter-images is still clearly incompatible with
the hypothesis that all of the high-mass GWTC-1 events are lensed events with intrinsic masses below 15 $\Msun$. 

\section{\label{sec:discussion}Conclusion and outlook}

We have performed the first focused search for strongly lensed counterpart images to
all binary black hole mergers from the GWTC-1 catalog~\cite{LIGOScientific:2018mvr}.
We recovered several candidates
previously found by \cite{Venumadhav:2019lyq,Nitz:2019hdf}.
Performing Bayesian inference on these candidate events
we found no evidence that these are lensed counterparts.
All other new candidates are consistent with a noise-only background.

The absence of clear candidates from our search
already provides useful observational constraints
on astrophysical lensing scenarios.
For example, if all observed \bbh[s] with component masses greater
than $15\,\Msun$ had originated from lower mass,
highly magnified, high redshift sources~\cite{Broadhurst:2018saj},
then we should have observed at least one counterpart.
We therefore rule out this hypothesis.

Another method to search for sub-threshold lensed events has been proposed in~\cite{Li:2019osa}.
Our single-template searches provide less freedom to match detector
noise fluctuations than the template bank employed in~\cite{Li:2019osa}.
For future applications,
the optimal template bank size per event might lie between our single-template method
and the larger banks of~\cite{Li:2019osa}.
For example, one could construct a template bank to obtain a certain minimal
match across each event's posterior distribution.
We will explore this in more detail in future work.

As discussed in Sec.~\ref{sec:setup}, our initial matched-filter search stage
currently does not check for
consistency in the extrinsic parameters (especially the sky location)
between any new candidates and the original event used to provide the search template.
This is a key area where the search could be improved in the future.
In order to include the extrinsic parameters of the signal a multi-detector coherent search \cite{Harry:2010fr}
could be used. This would improve the sensitivity of the search by removing candidates that
match well in the intrinsic parameters but have poor overlap in sky position.

Looking ahead, the LIGO-Virgo O3 run
has already yielded a rich crop of additional \gw candidates~\cite{gracedb},
and future observing runs promise many more~\cite{Aasi:2013wya}.
It has also been suggested that the first
detection of a lensed source is expected within the next
5 years~\cite{Ng:2017yiu,Li:2018prc}.
The framework of targeted sub-threshold searches
for lensed counterparts
as presented in this paper
can be readily applied to new detections in O3 and beyond.
Observing strongly lensed \bbh[s] before the detector network reaches design sensitivity \cite{Aasi:2013wya}
would imply that the merger rate increases much more steeply with redshift
than expected~\cite{LIGOScientific:2018jsj,Oguri:2018muv,Fishbach:2018edt},
or challenge the established understanding of lensing statistics.

More generally, once strongly lensed pairs of events can be identified,
joint parameter inference on the combined images
(as introduced in this paper and independently by~\cite{UWMPREP, RicoPREP})
can significantly improve estimates of the source properties and location~\cite{10.1093/mnras/staa2577}.
Hence, our approach is a key contribution towards the discovery of lensed GWs,
which will advance our understanding of both cosmology and high redshift black hole populations.

\vspace*{1cm}

Supplementary data for this paper is available at: \url{https://github.com/icg-gravwaves/lensed-o1-o2-data-release}

\begin{acknowledgments}
We thank Tjonnie Li, Otto Hannuksela, Rico Ka-Lok Lo and Jolien Creighton
for useful discussions.
IH is supported by STFC grant ST/T000333/1.
DK is supported by
the Spanish Ministry of Science, Innovation and Universities (ref. BEAGAL 18/00148)
and cofinanced by the Universitat de les Illes Balears,
as well as by grants from
the Ministry of Science, Innovation and Universities and
the Spanish Agencia Estatal de Investigación
(FPA2016-76821-P,    
RED2018-102661-T,    
RED2018-102573-E,    
FPA2017-90687-REDC,  
PID2019-106416GB-I00/AEI/10.13039/501100011033); 
the Vicepresid\`encia i Conselleria d’Innovaci\'o, Recerca i Turisme
and Conselleria d'Educaci\'o i Universitats del Govern de les Illes Balears;
the Comunitat Autonoma de les Illes Balears through the Direcci\'o General de
Pol\'itica Universitaria i Recerca with funds from the Tourist Stay Tax Law
ITS 2017-006 (PRD2018/24);
the Generalitat Valenciana (PROMETEO/2019/071);
the Fons Social Europeu;
European Union FEDER funds;
EU COST Actions CA18108, CA17137, CA16214, CA16104;
and the Spanish Ministry of Education, Culture and Sport grants
FPU15/03344 and FPU15/01319.
TEC is funded by a Royal Astronomical Society Research Fellowship.
This research has made use of data obtained from the Gravitational Wave Open Science Center,
a service of LIGO Laboratory, the LIGO Scientific Collaboration and the Virgo Collaboration.
LIGO is funded by the U.S. National Science Foundation.
Virgo is funded by the French Centre National de Recherche Scientifique (CNRS),
the Italian Istituto Nazionale della Fisica Nucleare (INFN)
and the Dutch Nikhef,
with contributions by Polish and Hungarian institutes.
The authors are grateful for computational resources provided by
the LIGO Laboratory
and supported by
National Science Foundation Grants PHY-0757058 and PHY-0823459,
as well as for additional computational resources
provided by Cardiff University
and funded by STFC grant ST/I006285/1.
This paper has been assigned document number \dcc.
\end{acknowledgments}

\bibliography{biblio}

\appendix

\section*{\supplement}
  
\squeezetable
\begin{table*}[t]
\setlength{\tabcolsep}{4pt}
\begin{tabular}{lccccc}
\toprule
delay-weighted $p$-value & {false-alarm rate$^{{-1}}$ [yr]} & UTC time & known event? & found by & {$|\Delta t|$ [d]} \\[0.5\baselineskip]
\midrule
$\ll$ 0.001 & $3.26\times10^{7}$ & 2017-08-14 10:30:43.5 & GW170814 & GW170104 & 222       \\
$\ll$ 0.001 & $3.26\times10^{7}$ & 2017-01-04 10:11:58.6 & GW170104 & GW170814 & 222       \\
$\ll$ 0.001 & $8.15\times10^{6}$ & 2017-08-23 13:13:58.5 & GW170823 & GW170818 & 5.5       \\
$\ll$ 0.001 & $3.62\times10^{6}$ & 2017-08-23 13:13:58.5 & GW170823 & GW170729 & 25        \\
$\ll$ 0.001 & $1.09\times10^{6}$ & 2017-01-04 10:11:58.6 & GW170104 & GW170809 & 217       \\
$\ll$ 0.001 & $1.78\times10^{3}$ & 2017-08-09 08:28:21.8 & GW170809 & GW170814 & 5.1       \\
$\ll$ 0.001 & $4.89\times10^{3}$ & 2017-08-09 08:28:21.8 & GW170809 & GW170818 & 8.7       \\
0.002 & 151       & 2017-08-14 10:30:43.5 & GW170814 & GW170809 & 5.1       \\
0.0035 & $3.94\times10^{3}$ & 2017-01-04 10:11:58.6 & GW170104 & GW170818 & 226       \\
0.0067 & $2.25\times10^{3}$ & 2017-01-21 21:25:36.6 & GW170121 & GW170818 & 208       \\
0.0075 & $1.93\times10^{3}$ & 2017-08-09 08:28:21.8 & GW170809 & GW170104 & 217       \\
0.0082 & 30        & 2017-08-14 10:30:43.5 & GW170814 & GW170818 & 3.7       \\
0.011 & 74        & 2017-08-23 13:13:58.5 & GW170823 & GW170809 & 14        \\
0.031 & 43        & 2017-07-29 18:56:29.3 & GW170729 & GW170823 & 25        \\
0.064 & 191       & 2017-01-21 21:25:36.6 & GW170121 & GW170823 & 214       \\
0.097 & 1.6       & 2017-07-27 01:04:30.0 & GW170727 & GW170729 & 2.7       \\
0.14 & 5.3       & 2017-08-09 08:28:21.8 & GW170809 & GW170823 & 14        \\
0.14 & 11        & 2015-09-14 09:50:45.4 & GW150914 & GW151012 & 28        \\
0.16 & 0.17      & 2017-07-30 08:05:26.8 &          & GW170729 & 0.55      \\
0.17 & 30        & 2017-08-23 13:13:58.5 & GW170823 & GW170814 & 9.1       \\
0.29 & $9.13\times10^{-5}$ & 2017-01-04 10:12:57.9 &          & GW170104 & $6.87\times10^{-4}$ \\
0.37 & 0.50      & 2017-08-04 14:57:29.3 &          & GW170809 & 4.7       \\
0.37 & 16        & 2017-03-04 16:37:53.4 & GW170304 & GW170729 & 147       \\
0.4 & $5.50\times10^{-4}$ & 2017-07-29 19:05:05.9 &          & GW170729 & $5.98\times10^{-3}$ \\
0.44 & 19        & 2017-01-04 10:11:58.6 & GW170104 & GW170823 & 231       \\
0.46 & $4.65\times10^{-4}$ & 2015-09-14 10:04:34.7 &          & GW150914 & $9.60\times10^{-3}$ \\
0.48 & 12        & 2017-03-04 16:37:53.4 & GW170304 & GW170823 & 172       \\
0.48 & $2.41\times10^{-4}$ & 2017-07-29 18:51:31.4 &          & GW170729 & $3.45\times10^{-3}$ \\
0.48 & $1.31\times10^{-4}$ & 2017-01-04 10:14:56.3 &          & GW170104 & $2.06\times10^{-3}$ \\
0.86 & 4.1       & 2017-08-23 13:13:58.5 & GW170823 & GW170104 & 231       \\
0.86 & 2.5       & 2017-04-01 08:19:53.0 &          & GW170809 & 13        \\
0.91 & 3.3       & 2017-01-04 10:11:58.6 & GW170104 & GW170729 & 206       \\
0.99 & 1.1       & 2017-03-04 16:37:53.4 & GW170304 & GW170818 & 166       \\

\bottomrule
\end{tabular}
 \caption{\label{table:candidates-all}
  List of all candidates with a delay-weighted \pval
  (first column)
  $< 0.5$ \emph{or}
  an inverse false-alarm rate
  (second column)
  of more than one year.
  In contrast to Table~\ref{table:candidates},
  all new candidates as well as any
  known events from both GWTC-1 and \cite{Venumadhav:2019lyq}
  are combined together, all sorted by the delay-weighted \pval.
  The third column gives the candidate end time (UTC)
  and the fourth column notes if an event at this time
  has already been published.
  The GWTC-1 event whose search found the reported candidate
  is listed in the fifth column,
  and the absolute value of the time delay between the two
  is given in the final column.
  }
\end{table*}

Table~\ref{table:candidates} in the main part of the paper
provides only those candidates
(found with
a delay-weighted \mbox{\pval $< 0.5$} or
an inverse false-alarm rate of more than one year)
which do not themselves correspond to another GWTC-1 event.
An extended version of the full search results
is provided here as Table~\ref{table:candidates-all},
which includes all candidates
above either of those two thresholds
from any of the ten searches,
and both sets of known events
from GWTC-1 and from \cite{Venumadhav:2019lyq}
are mixed in with the new candidates.

We also provide the full set of search results,
without thresholds on \pval or false-alarm rate,
as machine-readable supplementary data files.

\end{document}